\documentclass[structabstract]{aa} 
\usepackage{txfonts} \usepackage{graphicx}

\begin{document}

\title{Amplitude variability and multiple frequencies in 44 Tau:\\  2000 - 2006}

\author{M.~Breger \and P.~ Lenz}

\institute{Institut f\"ur Astronomie der Universit\"at Wien, T\"urkenschanzstr. 17,
A--1180 Wien, Austria\\
e-mail: michel.breger@univie.ac.at}

\date{Received date; accepted date}

\abstract
{}
{This study has three principal aims: (i) to increase the number of detected pulsation modes of 44 Tau, especially outside the
previously known frequency ranges, (ii) to study the amplitude variability and its systematics, and (iii) to examine the combination frequencies.}
 {During the 2004/5 and 2005/6 observing seasons, high-precision photometry was
 obtained with the Vienna Automatic Photoelectric Telescope in Arizona during
 52 nights. Together with previous campaigns, a data base from 2000 to 2006 was available for multifrequency analyses.}
 {Forty-nine pulsation frequencies  are detected, of which 15 are independent pulsation modes and 34 combination frequencies or harmonics.
The newly found gravity mode at 5.30 cd$^{-1}$ extends the known frequency range of instability. Strong amplitude variability
from year to year is found for the $\ell$ = 1 modes, while the two radial modes have essentially constant amplitudes. Possible origins of the amplitude
variability of the  $\ell$ = 1 modes, such as precession of the pulsation axis, beating and resonance effects are considered.
The amplitudes of the combination frequencies, $f_i + f_j$,  mirror the variations in the parent modes. The combination parameter, which relates the amplitudes of the
combination frequencies to those of the parent modes, is found to be different for different parents. 
}
   {}


\maketitle

\section{Introduction}

The asteroseismic comparison between measurements and theoretical models of the radial and nonradial pulsation of stars on and near the main sequence involves the exact matching of the values of the pulsation frequencies, as well as reproducing the frequency range in which instability occurs.
However, it is observationally difficult to determine the important borders of the frequency range in which a particular star is unstable. Due to the small amplitude growth rates of modes on the edges of the frequency instability range, only modes with small photometric amplitudes are expected to appear in these frequency regions.

If we consider the $\delta$ Scuti star FG Vir as an example (Breger et al. 2005), all photometric amplitudes in the 30 - 45 cd$^{-1}$ region had amplitudes less than 1.0 millimag. In fact, only the addition of 218 additional nights of photometry to the previous data allowed the detection of the pulsation modes near 40 cd$^{-1}$. In addition to the problem of small amplitudes in specific frequency regions, it is also essential to separate the `real' modes from the combination frequencies, $f_i \pm f_j$, which occur in these high-frequency and low-frequency regions. This separation requires very high frequency resolution so that the chance of accidental agreements between the expected and measured frequency values is essentially reduced to zero. In practice this requires campaigns covering several years.

The requirement of multiyear campaigns is strengthened by the amplitudes of many pulsators showing variability on timescales of months and, especially, years. To determine what types of modes show amplitude variability and to search for the astrophysical origin of this amplitude (and frequency) variability requires extensive observing campaigns. Finally, successful asteroseismology also demands the identification of individual pulsation modes. Both spectroscopic and photometric techniques have been applied: however, the large amounts of observing time required are easier to obtain on smaller telescopes with photometers.

The $\delta$ Scuti star 44 Tau is ideal for a detailed study because of its extremely low rotational velocity of 3 $\pm$ 2 km s$^{-1}$ (Zima et al. 2007), for which second-order effects of rotation can be neglected, and because its amplitudes place the star between the high-amplitude pulsators (HADS) with dominant radial modes and the average low-amplitude $\delta$ Scuti variable with mostly nonradial modes. Zima et al. (2007) present high-resolution spectroscopy of 44 Tau, which was used together with radial velocity data to derive the photospheric element abundances and provide $m$ values for the pulsation mode identifications. Six axisymmetric and two prograde modes were identified. 

In their asteroseismic study of 44~Tau, Lenz et al. (2008) present spherical mode degree identifications for the dominant modes of 44 Tau. Photometric amplitude ratios and phase differences suggest two radial, four $\ell$ = 1, and three $\ell = 2$ modes. Due to the measured $\log g$ value of 3.6 $\pm$ 0.1, both main sequence and post-main sequence models were examined. 
The identified modes can be fitted well in both evolutionary stages. The predicted frequency ranges for unstable modes are in good agreement with the observed ranges adopted by Lenz et al. (2008). Due to the different envelope structure of the main sequence and post-main sequence model, the borders of instability differ slightly. Consequently, new independent frequencies in the low-frequency and high-frequency regions possess important information for improving the models.

We have previously presented 90 nights of photometric observations obtained during the three observing seasons 2000/2001, 2001/2002, and 2002/3 (Antoci et al. 2007: this paper also lists a detailed observational history of the star). The data from three years led to detecting 29 frequencies, of which 13 were independent pulsation modes. The two comparison stars used showed low-frequency variability at the millimag level, which is, regrettably, quite common. Otherwise, these measurements are very accurate. Consequently, the publication omitted the 0 to 5 cd$^{-1}$ region awaiting future measurements with different comparison stars, as reported in the present paper.

\section{New photometry obtained 2004/5 and 2005/6}

The APT measurements were obtained with the T6 0.75~m Vienna Automatic Photoelectric Telescope
(APT), situated at Washington Camp in Arizona (Strassmeier et al. 1997, Breger \& Hiesberger 1999).
The telescope has been used before for several lengthy campaigns of the Delta Scuti Network, which confirmed
the longterm stability and millimag precision of the APT photometry. The $v$ and $y$ filters of the Str\"omgren system were used.
Only high-quality data were retained with the following number of nights and hours: 2004/5 observing season - 143 hours of photometry during 26 nights,
2005/6 observing season - 173 hours also during 26 nights.

The rapid position changes of the APT made it possible to observe 4 comparison stars:
HD 23626 (called C1, spectral type G0), HD 24301 (called C2, G0IV), and the two stars used during previous
campaigns HD 25867 (F1V) and HD 25768 (F8). The new comparison stars confirmed the suspicion that the two previously used comparison stars are slightly variable with periods longer than 1 day,
which would be typical of Gamma Doradus stars found at these spectral types. The results were already demonstrated in Fig. 3 of Breger (2007) as a warning to observers about common
millimag variables. These results also presented evidence for 0.936 and 0.97 cd$^{-1}$ frequencies in  HD~25867 and 0.885 cd$^{-1}$ in HD 25768 with millimag amplitudes.
The results allowed us to rereduce the 2000/1, 2001/2, and 2002/3 data by
prewhitening these frequencies from the comparison-star data. No variability was detected for the
two new comparison stars and the 2004/5 and 2005/6 measurements were reduced by
interpolating between these comparison stars. The comparison stars agreed to $\pm$ 2 mmag per single measurement in both the $y$ and $v$ passbands.

The new light curves with the final fits are not shown, since the diagrams look similar to the 2000 - 2003 measurements of 44 Tau published by Antoci et al. (2007).
The agreement between the observations and fits are excellent with residuals per single measurement of $\pm$ 2 mmag in the $y$ and $\pm$ 3 mmag in the $v$ passbands.

\section{Frequency analyses}

\begin{table*}[htb!]
\caption{Frequencies and amplitudes in the Str\"{o}mgren $v$ and $y$ filters.}
\label{table1}
\footnotesize
\begin{center}
\begin{tabular}{lcccccccccccc}
\noalign{\smallskip}
\hline\hline
\noalign{\smallskip}
& Frequency & Mode&\multicolumn{10}{c}{Amplitude in millimag}\\
 &cd$^{-1}$& $(\ell, m)$&\multicolumn{2}{c}{2000/1}&\multicolumn{2}{c}{2001/2}&\multicolumn{2}{c}{2002/3}&\multicolumn{2}{c}{2004/5}&\multicolumn{2}{c}{2005/6} \\
 &&or ID&$v$&y&$v$&y&$v$&y&$v$&y&$v$&y\\
\noalign{\smallskip}
\hline
\noalign{\smallskip}
$f_{1	}$ &	6.8980	&  (0,0)& $\it{39.64}$	&	27.37	&$\it{39.41}$&	27.46	&$\it{	39.32}$&	26.94	&	$\it{39.23}$&	27.42	&	$\it{39.53}$&	27.40	\\
$f_{2	}$ &	7.0060	&  (1,1) &$\it{19.11}$	&	13.30	&$\it{	16.80}$&	11.86	&$\it{	13.86}$&	9.19	&$\it{	10.00}$&	6.92	&	$\it{8.29}$&	5.64	\\
$f_{3	}$ &	9.1174	&  (1,1)& $\it{16.76}$	&	11.48	&$\it{	21.09}$&	14.56	&$\it{	17.62}$&	11.91	&$\it{	4.87}$&	3.37	&	$\it{0.81}$&	0.61	\\
$f_{4} $    &	11.5196	& (1,0)&	$\it{18.16}$	&	12.73	&$\it{	16.53}$&	11.65	&$\it{	16.72}$&	11.70	&$\it{	10.25}$&	7.21	&	$\it{8.78}$&	6.36	\\
$f_{5} $ 	&	8.9606	& (0,0)&	$\it{13.81}$	&	9.53	&$\it{	13.75}$&	9.24	&$\it{	12.82}$&	8.78	&$\it{	13.58}$&	9.49	&	$\it{13.77}$&	9.51	\\
$f_{6}$    &	9.5611	& (1,?)& $\it{	10.68}$	&	7.28	&$\it{	18.95}$&	12.87	&$\it{21.31:}$&	14.54:	&$\it{	0.61}$&	0.75	&	$\it{10.45}$&	7.07	\\
$f_{7	}$ &	7.3031	&  (2,0)&$\it{	6.67}$	&	4.54	&$\it{	6.11}$&	3.98	&$\it{	7.43}$&	4.91	&$\it{	9.00}$ & 6.45&	$\it{5.71}$ &	3.82	\\
$f_{8	}$ &	6.7955	&  (2,0)&$\it{	3.16}$	&	2.31	&$\it{	4.46}$&	3.02	&$\it{	2.55}$&	2.17	&$\it{	3.56	}$ & 2.37&	$\it{4.00}$&	3.05	\\
$f_{9	}$ &	9.5828	&  &$\it{	2.08}$	&	1.34	&$\it{	4.04}$&	2.36	&$\it{	4.70:}$&	3.23:	&$\it{	2.13	}$ & 1.47&	$\it{0.93}$&	0.79	\\
$f_{10	}$ &	6.3390	&  (2,?)&$\it{	2.34}$	&	1.70	&$\it{	2.15}$&	1.61	&$\it{	3.34}$&	1.85	&$\it{	1.93}$ &	1.30&	$\it{2.74}$&	2.00	\\
$f_{11	}$ &	8.6391	&  (?,0)&$\it{	2.17}$	&	1.60	&$\it{	1.89}$&	1.46	&$\it{	2.34}$&	1.66	&$\it{	3.24}$ &	2.19&	$\it{	2.91}$&	2.00	\\
$f_{12	}$ &	11.2947	& 		& &	0.62	& &	1.00	& &	1.21	& &	1.11	& &	1.19	\\					
$f_{13	}$ &	12.6915	&		& &	0.43	& &	0.32	& &	0.41	& &	0.09	& &	0.13	\\					
$f_{14	}$ &	5.3047	&		& &	0.59	& &	0.73	& &	1.04	& &	0.75	& &	0.41	\\					
$f_{15	}$ &	7.7897	&		& &	0.50	& &	0.77	& &	1.67	& &	0.28	& &	1.03	\\					
\multicolumn{3}{l}{Harmonics}\\
$f_{16	}$ &	13.7960	& 	=2$f_1$	& &	1.08	& &	1.11 & & 0.77 & &1.12 & & 0.80\\
$f_{17	}$ &	14.0120	& 	=2$f_2$	& &	0.69	& &	0.33	& &	0.60	& &	0.19	& &	0.31	\\
$f_{18	}$ &	23.0393	&	=2$f_4$	& &	0.78	& &	0.83	& &	0.71	& &	0.23	& &	0.25	\\
$f_{19	}$ &	17.9212	&	=2$f_5$	& &	0.25	& &	0.15	& &	0.16	& &	0.24	& &	0.18	\\
\multicolumn{3}{l}{Combination frequencies}\\
$f_{20	}$ &	13.9040	&	=$f_1$+$f_2$	& &	0.95	& &	0.83	& &	0.86	& &	0.49	& &	0.63	\\
$f_{21	}$ &	14.2011	&	=$f_1$+$f_7$	& &	0.24	& &	0.18	& &	0.37	& &	0.32	& &	0.21	\\
$f_{22	}$ &	15.8586	&	=$f_1$+$f_5$	& &	0.74	& &	0.93	& &	0.85	& &	0.49	& &	0.66	\\
$f_{23	}$ &	15.9666	&	=$f_2$+$f_5$	& &	0.33	& &	0.40	& &	0.30	& &	0.33	& &	0.27	\\
$f_{24	}$ &	16.0155	&	=$f_1$+$f_3$	& &	0.91	& &	1.02	& &	0.95	& &	0.31	& &	0.09	\\
$f_{25	}$ &	16.1234	&	=$f_2$+$f_3$	& &	0.33	& &	0.52	& &	0.27	& &	0.40	& &	0.40	\\
$f_{26	}$ &	16.4591	&	=$f_1$+$f_6$	& &	0.28	& &	0.83	& &	1.20	& &	0.17	& &	0.44	\\
$f_{27	}$ &	16.5671	&	=$f_2$+$f_6$	& &	0.33	& &	0.27	& &	0.45	& &	0.41	& &	0.32	\\
$f_{28	}$ &	18.3151	&	=$f_4$+$f_8$	& &	0.62	& &	0.78	& &	0.38	& &	0.26	& &	0.51	\\
$f_{29	}$ &	18.4177	&	=$f_1$+$f_4$	& &	1.13	& &	1.26	& &	0.97	& &	0.65	& &	0.61	\\
$f_{30	}$ &	18.5217	&	=$f_5$+$f_6$	& &	0.61	& &	0.82	& &	0.22	& &	0.44	& &	0.24	\\
$f_{31	}$ &	18.5256	&	=$f_2$+$f_4$	& & ... & & ... & & ... & & ... & & ...\\
$f_{32	}$ &	18.8228	&	=$f_4$+$f_7$	& &	0.10	& &	0.02	& &	0.34	& &	0.38	& &	0.16	\\
$f_{33	}$ &	20.4803	&	=$f_4$+$f_5$	& &	1.10	& &	0.91	& &	0.95	& &	0.70	& &	0.53	\\
$f_{34	}$ &	20.6371	&	=$f_3$+$f_4$	& &	0.37	& &	0.57	& &	0.43	& &	0.07	& &	0.12	\\
$f_{35	}$ &	21.0807	&	=$f_4$+$f_6$	& &	0.55	& &	0.49	& &	0.51	& &	0.24	& &	0.28	\\
$f_{36	}$ &	0.4437	&	=$f_6$-$f_3$   & &	0.31	& &	0.85	& &	1.04	& &	0.61	& &	0.35	\\
$f_{37	}$ &	0.9220	&	=$f_6$-$f_{11}$& &	0.76	& &	0.38	& &	0.88	& &	0.78	& &	0.14	\\
$f_{38	}$ &	1.9585	&	=$f_4$-$f_6$	& &	0.35	& &	0.78	& &	0.39:	& &	0.27	& &	0.35	\\
$f_{39	}$ &	2.0626	&	=$f_5$-$f_1$	& &	0.30	& &	0.61	& &	0.35	& &	0.48	& &	0.35	\\
$f_{40	}$ &	2.1114	&	=$f_3$-$f_2$	& &	0.04	& &	0.65	& &	0.43	& &	0.57	& &	0.52	\\
$f_{41	}$ &	2.2194	&	=$f_3$-$f_1$	& &	0.31	& &	1.09	& &	0.49	& &	0.43	& &	0.22	\\
$f_{42	}$ &	2.5551	&	=$f_6$-$f_2$	& &	0.44	& &	1.08	& &	0.45	& &	0.61	& &	0.45	\\
$f_{43	}$ &	2.6631	&	=$f_6$-$f_1$	& &	0.44	& &	0.83	& &	0.68	& &	0.16	& &	0.09	\\
$f_{44	}$ &	4.5136	&	=$f_4$-$f_2$	& &	0.61	& &	0.57	& &	0.31	& &	0.24	& &	0.51	\\
$f_{45	}$ &	4.6216	&	=$f_4$-$f_1$	& &	0.60	& &	0.69	& &	1.10	& &	0.44	& &	0.43	\\
\multicolumn{3}{l}{Triple combinations}\\
$f_{46	}$ &	25.4236	&	=$f_1$+$f_2$+$f_4$& &	0.19	& &	0.19	& &	0.07	& &	0.15	& &	0.09	\\
$f_{47	}$ &	27.9788	&	=$f_1$+$f_4$+$f_6$	& &	0.23	& &	0.15	& &	0.28	& &	0.17	& &	0.09	\\
$f_{48	}$ &	29.9373	&	=$2f_4$+$f_1$	& &	0.42	& &	0.32	& &	0.16	& &	0.23	& &	0.08	\\
$f_{49	}$ &	31.9999	&	=$2f_4$+$f_5$	& &	0.25	& &	0.19	& &	0.10	& &	0.08	& &	0.08	\\
\noalign{\smallskip}
\hline
\end{tabular}
\begin{flushleft}
Note: Some close frequency pairs are near the annual frequency resolution.
Consequently, annual $f_{31}$ values are not given and for 2002/3, some amplitudes marked with a colon are uncertain.
\end{flushleft}
\end{center}
\end{table*}

The pulsation frequency analyses were performed with a package of computer
programs with single-frequency and multiple-frequency techniques (PERIOD04,
Lenz \& Breger 2005),
which utilize both Fourier and multiple-least-squares algorithms. The latter technique fits up to several
hundreds of simultaneous sinusoidal variations in the magnitude domain and does not rely
on sequential prewhitening. The amplitudes and phases of all modes/frequencies are determined by minimizing
the residuals between the measurements and the fit. The frequencies can also be improved at the same time.

New frequencies of pulsation were searched for by applying Fourier analyses to the data prewhitened by the previously
detected frequencies. However, due to strong amplitude variability from year to year, as well as the different
amplitudes and phases found in the $v$ and $y$ passbands, for each year separate multifrequency solutions were made, which could then be
prewhitened to search for additional frequencies. This approach works well for modes with amplitudes in excess of 1 mmag. For
the many modes with small amplitudes, we improved the signal-to-noise ratios by combining the data from the $v$ and $y$ passbands.
This was done by scaling the $v$ data by an experimentally determined factor of 0.69 and increasing the weights of the scaled
data accordingly. In principle, this could create problems due to small phase shifts, but this is negligible for modes with such small amplitudes.
Following Breger et al. (1993), a significance criterion of an amplitude signal-to-noise ratio of 4.0
(which corresponds to a power signal-to-noise ratio $\sim$12.6) was adopted for all frequencies.

The results are shown in Table 1. For $f_{12}$ to $f_{49}$, we only list the combined amplitude for the $v$ and $y$ passbands.
Due to the scaling of the $v$ data, this corresponds to the $y$ passband, whose effective wavelength is similar to that of the standard Johnson $V$ passband.
The pulsation mode identifications listed are from
Zima et al. (2007) and Lenz et al. (2008). The formal uncertainties in the amplitudes are listed in Table 2.
These values need to be considered in discussing the amplitude variability.

\begin{figure}
\centering
\includegraphics[bb=40 430 565 800,width=85mm,clip]{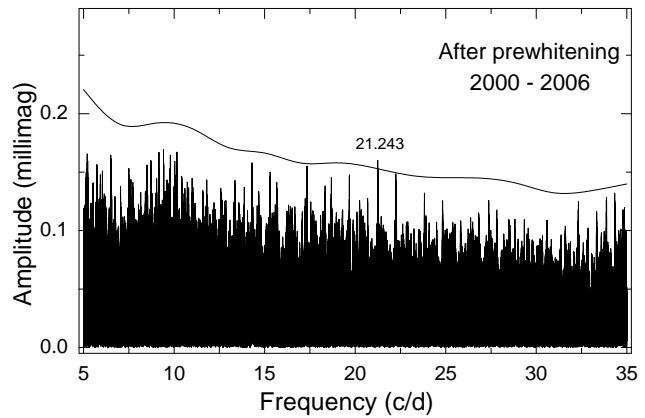}
\caption{Fourier spectra of the residuals in the 5 - 35 cd$^{-1}$ region after prewhitening 49 frequencies. The drawn curve represents the calculated
significance level.}
\end{figure}

\begin{figure}
\centering
\includegraphics[bb=37 150 565 800,width=85mm,clip]{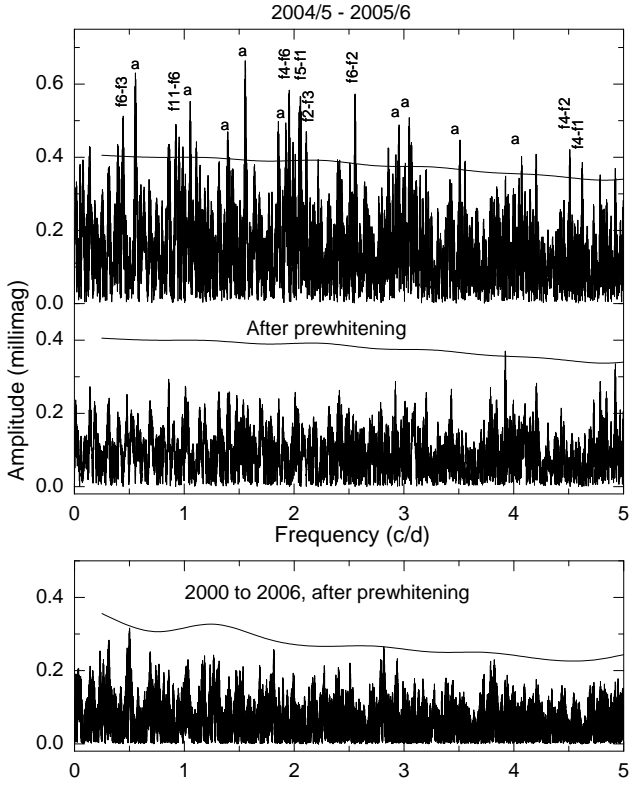}
\caption{Fourier spectra of the low-frequency region of 44 Tau. Note the many identified combination frequencies and their 1 cd$^{-1}$ aliases.
We have shown the Fourier spectrum of the final residuals for both the 2004/5 and 2005/6, which used excellent
comparison stars, and the total data set. The statistical significance levels are also shown.}
\end{figure}

\begin{figure}
\centering
\includegraphics[bb=40 250 570 800,width=85mm,clip]{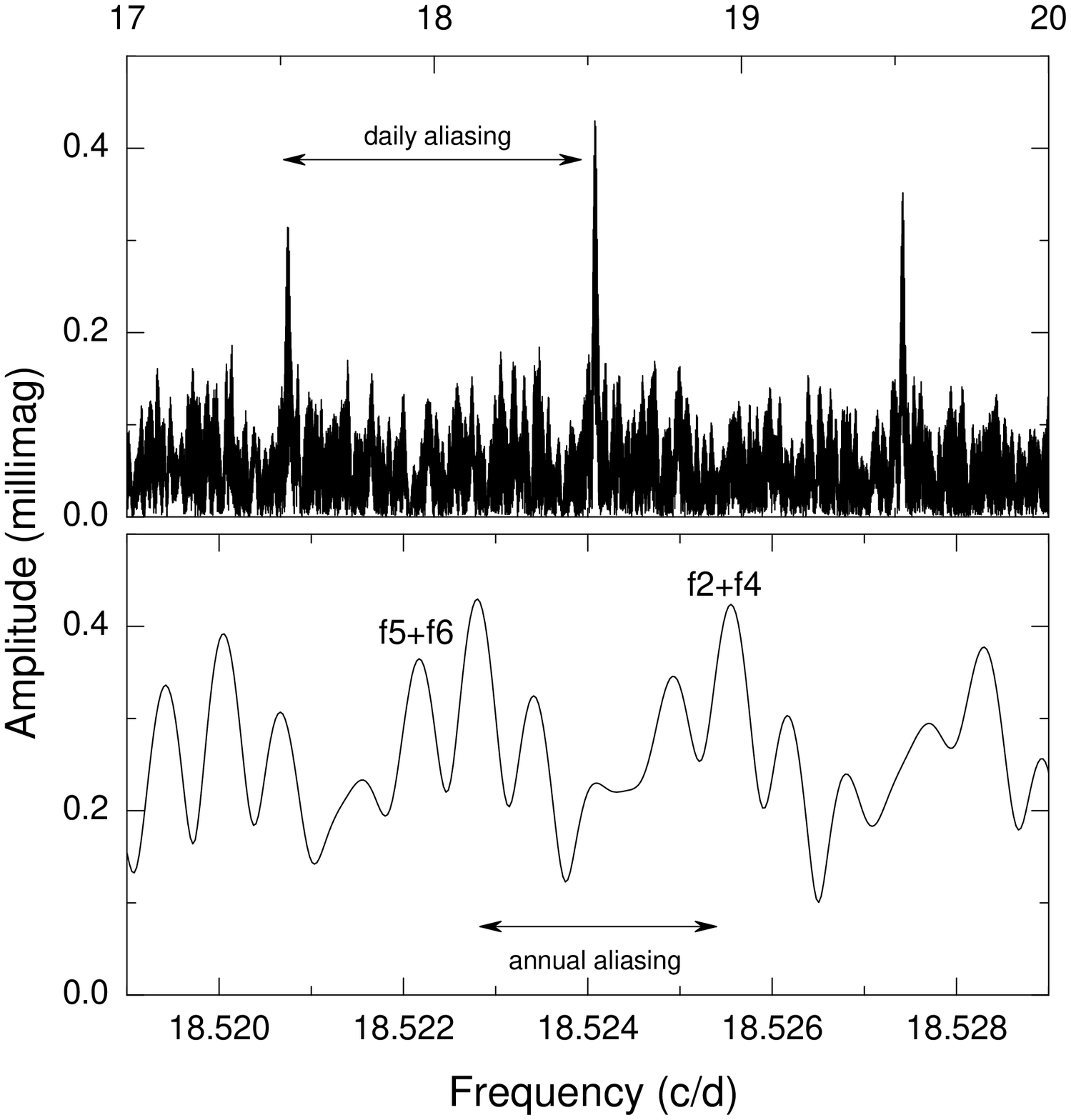}
\caption{Two combination frequencies with similar values. The diagram shows that the frequency resolution of the six-year data set is high enough to resolve
these two frequencies.}
\end{figure}

\subsection{The main pulsation region}

The main pulsation region, in which we find independent modes, ranges from 5.3 to 12.7 cd$^{-1}$, which is more than the previously known range.
Two previously unknown frequencies, representing independent pulsation modes, were discovered. These are found at 5.30 and 7.79 cd$^{-1}$, respectively. In both cases, the
amplitude signal-to-noise ratio is higher than 10, representing certain detections. The 7.79 cd$^{-1}$ mode is near
a 1 cd$^{-1}$ alias of the previously known 6.80 cd$^{-1}$ mode. However,
we found it impossible to identify the new frequency as an artifact of the 6.80 cd$^{-1}$ mode, caused by amplitude variability.

Are there additional modes seen in the data? Two more peaks are near the statistical detectability limit of the amplitude signal-to-noise ratio of 4.0: 
a peak at 21.243 cd$^{-1}$ has an amplitude near 0.15 mmag (Fig. 1), while the 3.92 cd$^{-1}$ peak
is seen only in the 2004/5 and 2005/6 data with a height of 0.4 mmag.
The difficulty does not lie with the signal, but with the noise figure. We do not regard the calculated noise as reliable;
in these frequency ranges, there are many correctly identified combination frequencies, which have been prewhitened. However, after each prewhitening, a zero
amplitude remains at the prewhitened frequency value, while there should actually be some noise with different phasing left. After prewhitening a number of these frequencies,
the overall noise is reduced. Evidence of this effect can be seen in Fig. 2, since the expected increase in noise toward lower frequencies is not seen. Consequently, modes near
the formal statistical detection limit should be treated with caution; i.e., we have not accepted the two peaks near the detection limit. An additional argument can be made that
such an uncertain detection should be seen in two or more independent data sets: this is not the case here. The two peaks may, therefore, be caused
by noise.

What mode does the new frequency at 5.30 cd$^{-1}$ represent? The mode at 5.30 cd$^{-1}$ increases the range of excited frequencies into the
g-mode domain, since the radial fundamental mode is found at 6.90 cd$^{-1}$. The identification of the 6.90 cd$^{-1}$ as the radial fundamental mode is secure because of photometric, as well as spectroscopic, mode identifications, the observed frequency ratio of 0.770 between the observed 6.90 and 8.96 cd$^{-1}$ modes and the known parallax. The frequency of the newly discovered 5.30 cd$^{-1}$ mode also gives a frequency ratio of 0.769 relative to the known radial fundamental mode: the near agreement with the radial ratio of 0.770 serves as a warning that the frequency ratio alone is not a sufficient criterion for mode identification.

\subsection{Combination frequencies}

The excellent frequency resolution provided by the coverage from 2000--2006 makes it possible to accurately identify those peaks that are combinations, $f_i\pm f_j$, as well
as $2f_i$ harmonics. We find that the peaks outside the 5.3 to 12.7 cd$^{-1}$ range can be identified as combinations or harmonics. We find it remarkable that even some triple combinations are seen. The parent modes with the highest amplitudes were involved most often in the combination frequencies, which is not surprising.

An interesting situation arises because of the similarity in some predicted frequency combinations. Numerically, the frequency values of three $\ell$ = 1 modes $f_2$, $f_4$, $f_6$, and the radial mode $f_5$ are related in the sense that

$f_2$ + $f_4$ = $f_5$ + $f_6$ + 0.0039 cd$^{-1}$.\\
This is a very close, probably accidental, agreement. The small frequency separation is only of the order of the aliasing
caused by the gaps between the annual observing seasons. For the whole time period of six years, the frequency resolution
is high enough to separate both combinations. Indeed, Fig. 3 shows that both are present, but not at equal amplitudes.
However, for a single observing season, the frequency resolution is too low to separate the two combination frequencies; consequently,
we only derive a combined amplitude.

A similar situation exists for combination frequencies formed by frequency differences $f_5$ - $f_2$ vs. $f_4$ - $f_6$ near 1.95 cd$^{-1}$ and
$f_6$ - $f_2$ and $f_4$ - $f_5$ near 2.56 cd$^{-1}$. However, in these cases the power spectrum covering five years only shows unique, single
detections.

\par
\begin{table}[htb!]
\caption{Formal uncertainties of the amplitude values in mmag}
\label{table2}
\footnotesize
\begin{center}
\begin{tabular}{lccc}
\noalign{\smallskip}
\hline\hline
\noalign{\smallskip}
Observing season&$\sigma(v)$&$\sigma(y)$&$\sigma$(combined $y$)\\
\noalign{\smallskip}
\hline
\noalign{\smallskip}
2000/1 & 0.07 & 0.06 & 0.04\\
2001/2  & 0.12 & 0.12 & 0.07\\
2002/3 & 0.19 & 0.15 & 0.10\\
2004/5 & 0.09 & 0.07 & 0.05\\
2005/6 & 0.08 & 0.07 & 0.05\\
\noalign{\smallskip}
\hline
\noalign{\smallskip}
\end{tabular}
\newline
Extensive Monte Carlo simulations give larger uncertainties\\ with a factor of $\sim$1.35.
\end{center}
\end{table}
\section{Amplitude variability: Observations}

\begin{figure*}
\centering
\includegraphics[bb=0 30 800 500,width=175mm,clip]{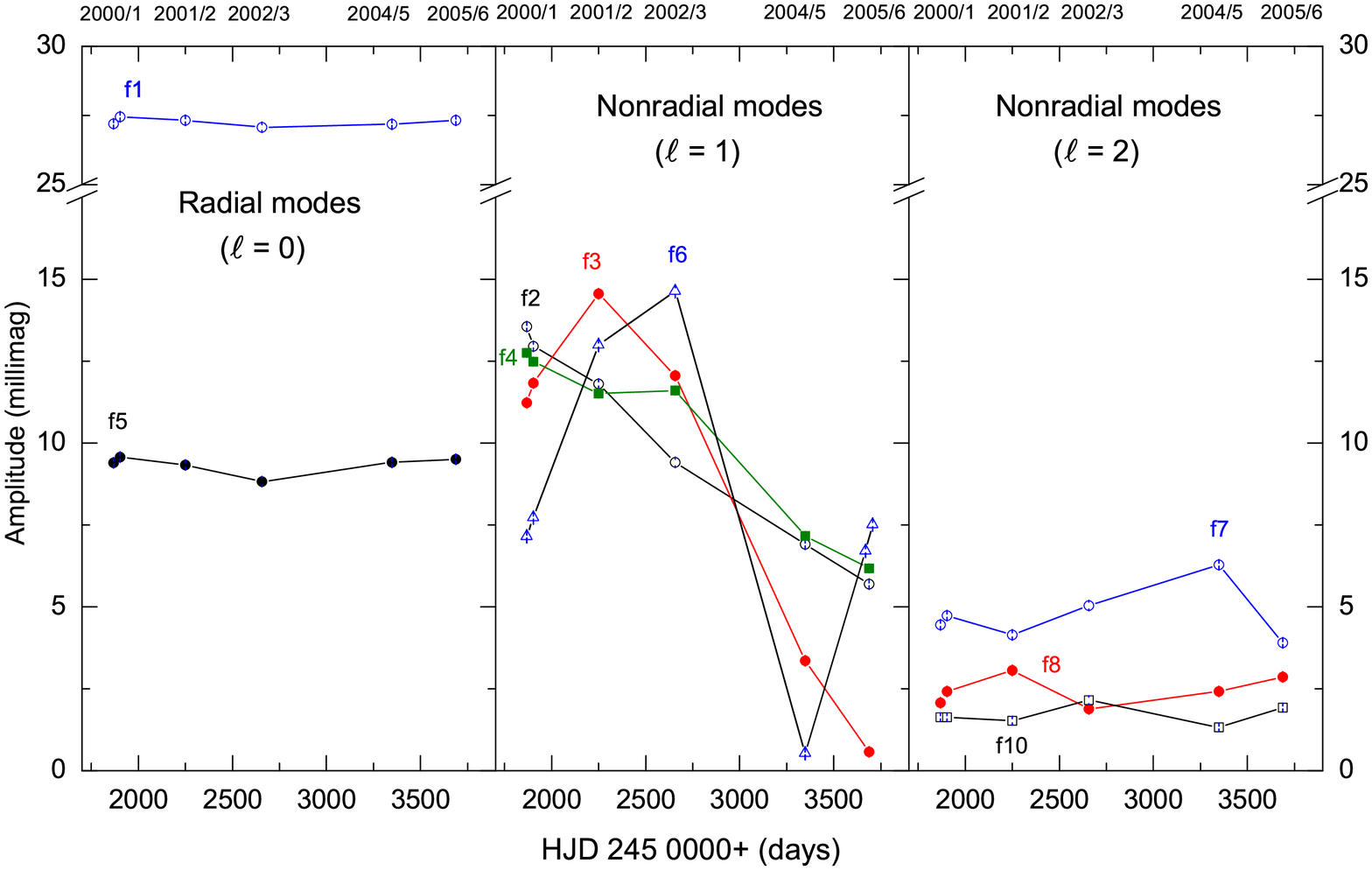}
\caption{Amplitude variability of radial and nonradial modes found in 44 Tau for the time period 2000-2006. The amplitudes in both the $y$ ($\sim V$) and the $v$ passbands were used after scaling the $v$ amplitudes. The error bars are shown and are usually the same size as the symbols used. Note the relative
constancy of the radial modes and the large variability of the $\ell$ = 1 modes. The variations within the 2000/1 and 2005/6 ($f_6$ only) observing seasons confirm the long-term trends.}
\label{fig:ampvar}
\end{figure*}

It has already been noticed by Poretti et al. (1992) that in 44 Tau the amplitudes of $f_2$ and $f_6$ are variable.
The new extensive data covering five observing seasons in six years allow us to examine these
and other variations in more detail. For the dominant modes with mode identifications, the amplitude variability is shown in Fig. 4.
Here the long 2000/1 observing season was subdivided into two parts with the average time strings spaced 40 days apart. Separate multifrequency
solutions show that the small variations are  as expected from the larger year-to-year variations and rapid changes are not seen. Note that the steady amplitude
increase in $f_6$ after 2004/5 is confirmed when the 2005/6 season is subdivided into two parts.

\begin{figure*}
\centering
\includegraphics[bb=37 320 800 800,width=175mm,clip]{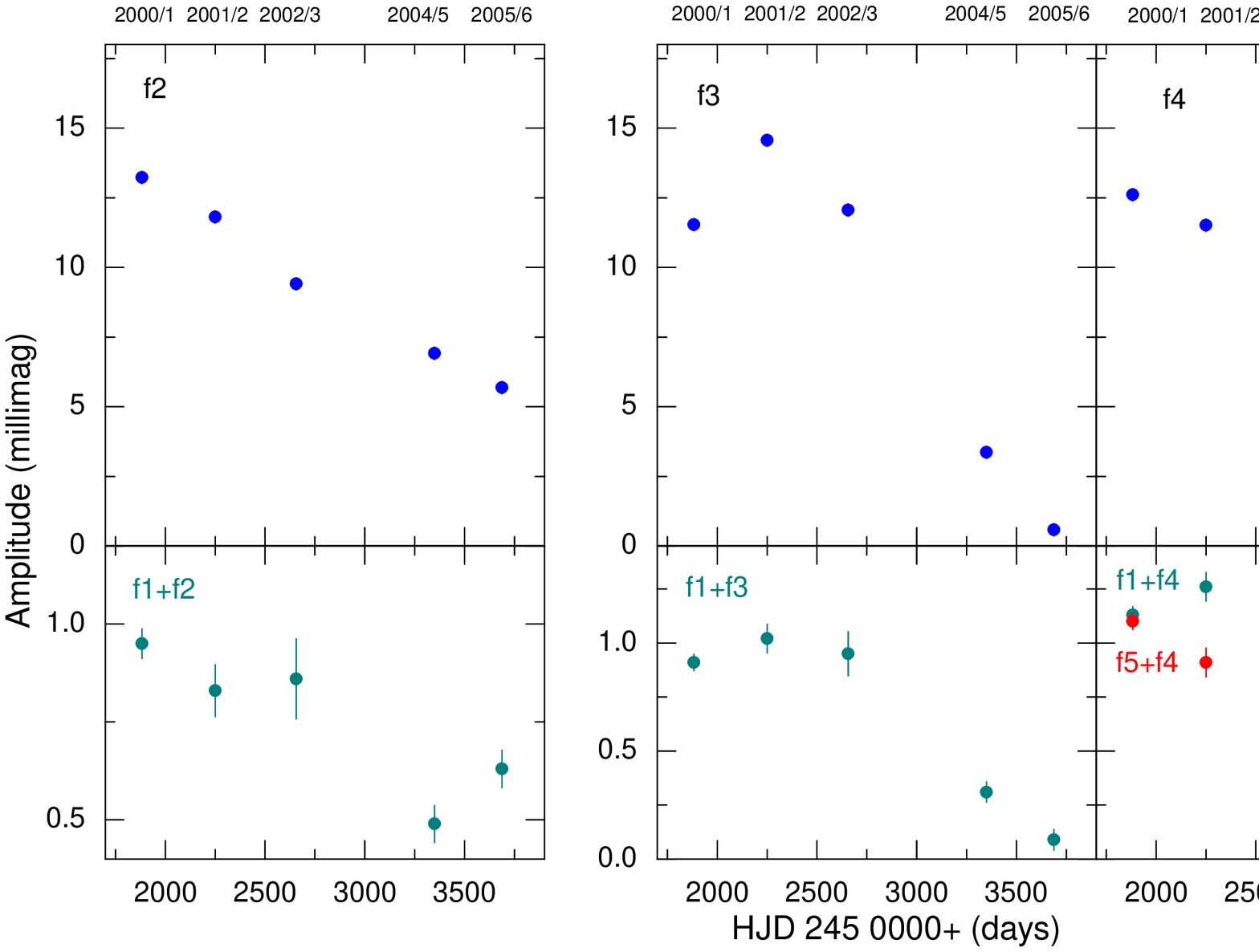}
\caption{The relation between amplitude variability of combination frequencies and their parents. We have selected only modes for which one of the parents is a radial mode with constant (or nearly constant) amplitudes. The diagram demonstrates the excellent correlations.}
\end{figure*}

The observed amplitude variability can be summarized as follows:

(i) The two radial modes at $f_1$ and $f_5$  show very little or no amplitude variability from 2000 to 2006. For both modes,
the new amplitude values also agree with the amplitudes from 1989/1990 given by Poretti et al. (1992). The results are consistent with amplitude constancy.

(ii) \emph{All} four $\ell$ = 1 modes ($f_2$, $f_3$, $f_4$, and $f_6$) show strong amplitude variability with decreasing amplitudes between 2000 and 2006.
The timescale of variation is several years with little variation seen within an observing season. A general trend of decreasing amplitudes from
2000 to 2006 exists. However, the different modes do not vary together: the differences in behavior of the four modes considerably exceed the observational uncertainties. The different behavior is not connected to the fact that the modes at $f_2$ and $f_3$ are prograde, while $f_4$ is axisymmetric.

(iii) The $\ell$ = 2 modes have smaller photometric amplitudes, but the amplitude variability exceeds the statistical uncertainties. Otherwise, no systematic behavior is seen.

\subsection{Amplitude variability of the combination frequencies}

The observed combination frequencies in $\delta$ Scuti stars may be created by two mechanisms. Resonant excitation of a pulsation mode close to the combination frequency leads to frequency synchronization at exactly $f_i + f_j$. The second possibility is that a nonlinear response of the medium in the outer layers of the star causes pulsation signals at combination frequencies. The latter mechanism is likely to be responsible for combination frequencies in DA and DB white dwarfs (see e.g. van Kerkwijk et al. 2000).

If the same mechanism is responsible for the amplitude variability in 44 Tau, we can expect the combination frequencies to mirror the amplitude variations of the parent modes.
The strong amplitude variability of the $\ell$ = 1 modes makes it possible to examine the amplitudes of the combination frequencies, $f_i \pm f_j$, involving one or two of these modes.
Three difficulties arise: if both $f_i$ and $f_j$ are variable, it becomes difficult to separate the effects from the two components.
This can be avoided if one of the parents of the combination frequencies has a constant amplitude; i.e., in 44 Tau one of the parents should be a radial mode.
Secondly, the parent modes need to have high amplitudes for a combination frequency to be visible. This restricts us to
the dominant modes. Thirdly, the amplitudes of the combination frequencies are small and affected by random or systematic noise; consequently, we have omitted the
low-frequency region. The combination frequencies shown in Fig. 5 obey the criteria and a very good correlation is seen.

The agreement between the amplitudes of the combination and the parent modes is very good. This argues against an origin of the combination frequencies as independent modes excited by resonance: with resonance the power can be redistributed between the three modes, but not just decreased. The measurements, however, match the effects due to nonlinear mixing of the eigenmodes, as discussed by Garrido \&  Rodr\'iguez (1996).

The amplitudes given in Table 1 are used to calculate the combination parameter, $\mu$, which
relates the amplitude of the combination frequency with those of the parents,

\begin{equation}
A_{\rm comb} = \mu \cdot A_i \cdot A_j,
\end{equation}
where $A_i$ and $A_j$ are the amplitudes of the parent modes. This definition is similar to the definition of van Kerkwijk et al. (2000). 
We have calculated the sizes of the combination parameter for the different combinations including a complete treatment of the propagation errors. The uncertainties
in the amplitudes (given in Table 2) were used after multiplying the values by a factor of 1.35 derived from
Monte Carlo simulations using the real data. The data from the different years were
averaged with weights derived from the known uncertainties of the annual amplitudes.
The resulting coefficients, $\mu$, are shown in Table~3. Regrettably, the coefficients of other interesting combinations are too uncertain to be listed.

\par
\begin{table}[htb!]
\caption{Relationship between the amplitudes of combination frequencies and parent modes.}
\footnotesize
\begin{center}
\begin{tabular}{ccc}
\noalign{\smallskip}
\hline\hline
\noalign{\smallskip}
\multicolumn{2}{c}{Combination frequencies}  &  Combination parameter\\
$f_i+f_j$ & ($\ell, m$) + ($\ell, m$) & $\mu$ = A$_{\rm comb}$/(A$_i$ $\cdot$ A$_j$)\\
\noalign{\smallskip}
\hline
\noalign{\smallskip}
$f_1+f_5$ & (0,0) + (0,0) & 0.0027 $\pm$ 0.0005\\
\noalign{\smallskip}
$f_1+f_2$ & (0,0) + (1,1) & 0.0028 $\pm$ 0.0004 \\
$f_1+f_3$ & (0,0) + (1,1) & 0.0028 $\pm$ 0.0002 \\
$f_1+f_4$ & (0,0) + (1,0) & 0.0034 $\pm$ 0.0003 \\
$f_5+f_4$ & (0,0) + (1,0) & 0.0092 $\pm$ 0.0005 \\
\noalign{\smallskip}
Uncertain values\\
\noalign{\smallskip}
$f_4+f_6$ & (1,0) + (1,?) & 0.0045 $\pm$ 0.0018\\
$f_5+f_2$ & (0,0) + (1,1) & 0.0071 $\pm$ 0.0019\\
\noalign{\smallskip}
\hline
\end{tabular}
\end{center}
\end{table}

One may assume that the combination frequency involving a radial and a $\ell$ = 1 mode shows the same geometrical cancellation effects as the $\ell = 1$ mode of the combination, while there is no light cancellation for the radial mode. Consequently, the combination parameters in Table 3 can be considered to already be free of geometrical cancellation effects.

We find that

(i) the combinations involving the radial fundamental mode, $f_1$, all have similar combination parameters near 0.003. This includes the
combinations with the $\ell$ = 1 and $\ell$ = 0 modes.

(ii) The two radial modes have different combination parameters; although the first overtone, $f_5$, has a considerably smaller amplitude
than the fundamental mode, $f_1$, the combination frequencies ($f_1$+$f_4$)  and ($f_5$+$f_4$) have similar amplitudes, A$_{\rm comb}$.

(iii) Some (weak) evidence exists that the combinations of the first radial overtone, $f_5$, with other nonradial modes also have higher values of the
combination parameter.

\section{Theoretical interpretation of the observed amplitude variability of the $\ell=1$ modes}

The exceptionally slow rotation of 44 Tau needs to be considered when analyzing amplitude variability. The reason for the slow rotation of 44 Tau has not yet been revealed.
Common reasons for slow rotation are magnetic braking, a close companion star, or planets. However, the observed spectra do not show any indications of a strong magnetic field or a binary star.  

We examined different scenarios to explain the observed amplitude variability in 44 Tau. Only for one of the four $\ell = 1$ modes, $f_6$, was a full modulation period presumably covered. Its duration was approximately 5 years. We now consider several mechanisms that cause amplitude modulation: beating of close frequencies, resonant coupling of modes, and/or the precession of the pulsation axis that causes a variation in the visibility of a pulsation mode.  

\subsection{Amplitude variability due to precession of the pulsation axis}

A possible scenario that causes amplitude variability for $\ell~=~1$ modes, but does not influence the photometric amplitudes of radial modes, is the precession of the pulsation axis, which is usually assumed to coincide with the rotation axis. The variation in the stellar inclination angle results in a changing cancellation of light for $\ell = 1$ modes, while the visibility of $\ell = 0$ modes remains the same.

Zima et al. (2007) have identified two $\ell$ = 1 modes, $f_2$ and $f_3$, as prograde modes ($m$ = 1) and $f_4$ as an axisymmetric mode ($m$~=~$0$). For the fourth $\ell=1$ mode, $f_6$, the azimuthal number could not be identified. The measured inclination in 2004 was 60~$\pm$~25$^{\rm o}$. We computed mode amplitudes in the Str\"omgren $y$ band according to the equations given in Daszy\'nska-Daszkiewicz et al. (2003) assuming the intrinsic mode amplitude, $\varepsilon$, to be 0.000125. This value was estimated by scaling the predicted photometric amplitudes to the observed amplitudes. The actual value for $\varepsilon$ is unimportant in this study because we only want to examine the effect of the inclination on mode visibility and not to predict absolute photometric amplitudes. Our computation also takes limb darkening effects into account. The frequency dependence between the relative gravity changes and the radius changes needs to be considered, and one has to consider the frequency of the given mode as well. However, the influence of the frequency value on the visibility is weak. 
In this examination we relied on the post-main sequence model presented in Lenz et al. (2008). It should be noted that the specific model parameters do not change the main results in this section significantly.

\begin{figure}
\centering
\includegraphics[bb=45 40 750 600,width=85mm,clip]{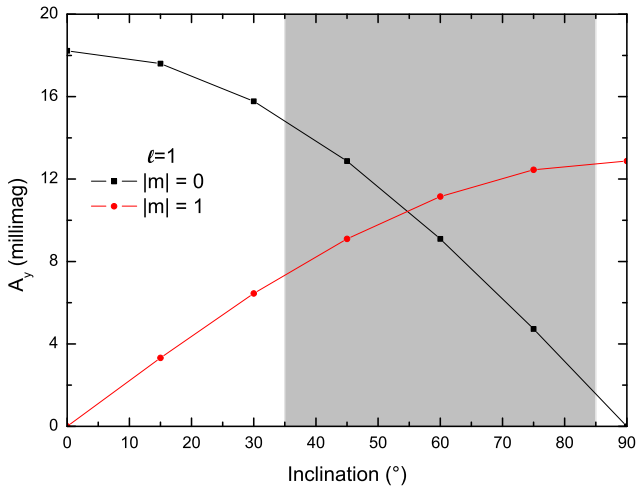}
\caption{Dependence of mode visibility in the Str\"omgren $y$ filter on the stellar inclination shown for an axisymmetric and a non-axisymmetric $\ell = 1$ mode with the same frequency. The shaded area marks the measured limits for the inclination of 44 Tau.}
\label{fig:inc}
\end{figure}

The dependence of photometric amplitudes on stellar inclination in the case of the $\ell = 1$ mode $f_6$ is shown in Fig. 6. It can be clearly seen that the visibility of an axisymmetric and non-axisymmetric mode is different. The intrinsic mode amplitudes are assumed to be constant, and the different visibility is only due to cancellation effects. The amplitude of a $(\ell,m) = (1,0)$ mode decreases with increasing inclination and finally is fully cancelled out at 90$^{\rm o}$ (equator-on view). The visibility of a $(\ell,m) = (1,1)$ mode is, however, highest for large inclinations and lowest at small inclinations (pole-on view).

\begin{figure}
\centering
\includegraphics[bb=40 40 750 550,width=85mm,clip]{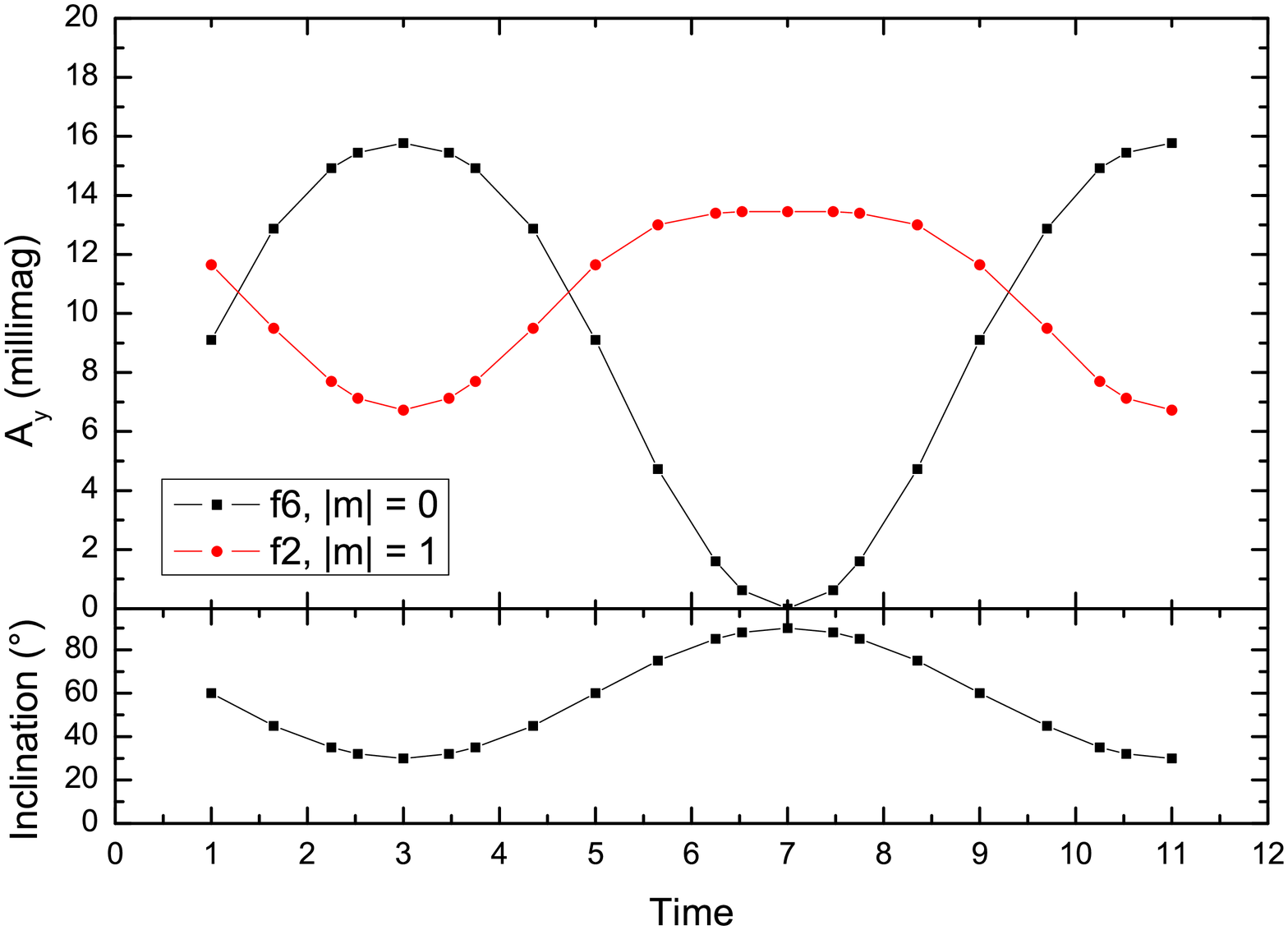}
\caption{Predicted amplitude variability due to precession of the pulsation symmetry axis for an inclination varying between 30 and 90$^{\rm o}$. Lower panel: assumed variation of the inclination in an arbitrary time unit. Upper panel: Str\"omgren $y$ amplitudes of the $(\ell,m) = (1,0)$ mode $f_6$ and the $(\ell,m) = (1,1)$ mode $f_2$ computed for the corresponding inclination values. }
\label{fig:prec}
\end{figure}

The effect of the precession of the pulsation axis on the visibility of the $\ell = 1$ modes $f_2$ and $f_6$ is shown in Fig.~\ref{fig:prec}. Here we assume that the inclination varies between 30 and 90$^{\rm o}$ and that $f_6$ is an axisymmetric mode. This diagram looks very similar when $f_3$ and $f_4$ are plotted. The predicted change of visibility of the axisymmetric $f_6$ mode mimics the observed amplitude variability very well. This would indicate an actual inclination angle close to 85$^{\rm o}$ for the year 2004. The inclination angle was previously determined in 2004 with a mean value of 60 $\pm$ 25$^{\rm o}$. Therefore, an inclination close to 85$^{\rm o}$ is still within the given uncertainty limit. However, the behavior of the non-axisymmetric mode $f_2$ is clearly different and its amplitude expected to be highest at large inclination angles. As can be seen from Fig.~\ref{fig:ampvar}, this is not observed. Furthermore, the second axisymmetric $\ell=1$ mode, $f_4$, does not go to zero amplitude simultaneously with $f_6$. Consequently, the precession of the stellar inclination cannot explain the observed amplitude variability of all the $\ell$ = 1 modes.

\subsection{Beating effects}

Beating effects between close frequencies often explain short-scale amplitude variability in $\delta$ Scuti stars (e.g., FG Vir, Breger \& Pamyatnykh 2006).
Since 44 Tau rotates at a very low rate (3~$\pm$~2 km s$^{-1}$), the rotationally split components are separated by only a very small amount.
In such a case the components of the multiplet may cause amplitude modulation if the frequencies are too close to be resolved in the observed data set.

\begin{figure}[htb]
  \centering
  \includegraphics[bb=45 35 750 600,width=85mm,clip]{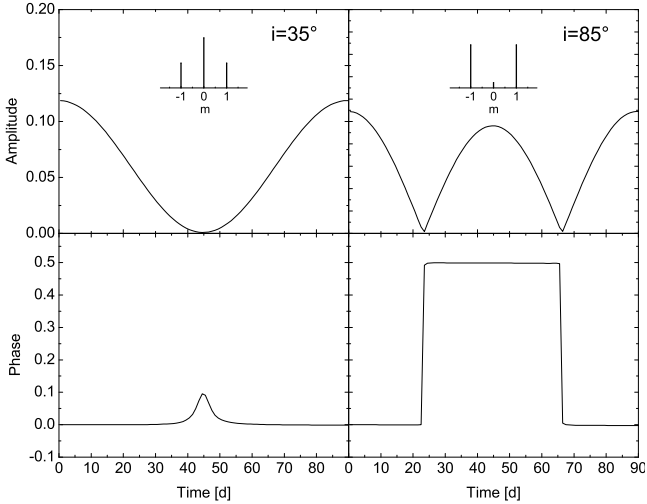}
  \caption{Expected amplitude and phase modulation for three-mode beating considering geometrical visibility conditions for an inclination angle of 35$^{\rm o}$ (left panels) and 85$^{\rm o}$ (right panels). The amplitudes of the components of the triplet are given in the inset diagrams. The rotational splitting was computed for the measured mean rotation rate of 3 km s$^{-1}$.  }
  \label{fig:splitting}
\end{figure}

Since we intend to examine this possibility for all observed $\ell$~=~1 modes in 44 Tau, we carried out tests with artificial data sets to determine the expected amplitude and phase modulation for three-mode beating. The differing visibility of the components of the multiplet depending on the inclination angle needs to be taken into account. Consequently, we examined the situation that occurs for the two boundary cases for the inclination of 44 Tau, $i$ = 35 and 85$^{\rm o}$. The inset diagram in Fig.~\ref{fig:splitting} shows the visibility of the multiplet components for these inclination angles. For an inclination of 85$^{\rm o}$, the observed amplitude of the central peak ($m=0$) is close to zero, and we essentially have a similar situation as for two-mode beating.

To examine the effect of three-mode beating, we computed the rotational splitting of the $\ell$ = 1 mode $f_6$ according to 2nd-order theory assuming a rotation rate of 3 km s$^{-1}$. The frequency separation of 0.011 cd$^{-1}$ is almost equidistant at such a low rotation. We created an artificial time string using the corresponding amplitudes for the given limits of the inclination angle. This data set was then analyzed with PERIOD04 assuming a single unresolved frequency centered at the axisymmetric mode. The amplitude and phase variability that occurs in such a case is given in Fig.~\ref{fig:splitting} for the inclination angles 35 and 85$^{\rm o}$, respectively.

An important observable is the occurrence of a phase change at minimum amplitude. The minimum of the amplitude modulation cycle has been observed only for $f_6$. No significant phase change was detected. However, the determination is uncertain because the exact time of the minimum is not known, and due to observing gaps, the uncertainties in the phase determination are large. Consequently, we can only draw conclusions from the expected timescales of the beating effect.

The period of amplitude modulation, $P_{\rm mod}$, corresponds to $1/\Delta \nu_{0,\pm1}$ where $\Delta \nu_{0,\pm1}$ is the mean frequency spacing between the $m = 0$ and $|m| = 1$ components. As already mentioned earlier, at an inclination angle of 90$^{\rm o}$, the observable amplitude of the axisymmetric mode vanishes and two-mode beating occurs with a modulation period $1/{\Delta \nu_{-1,+1}}$, where $\Delta \nu_{-1,+1}$ denotes the frequency separation between the two non-axisymmetric components of the triplet. Since the inclination of 44 Tau may be as high as 85$^{\rm o}$, we cannot exclude such a case.

These results allow us to examine the expected timescales of beating effects for 44 Tau. At the upper limit of the rotation rate, 5 km s$^{-1}$, the rotational splitting amounts to 0.0185~cd$^{-1}$, which causes amplitude modulation periods of 54 d for three-mode beating and 27 d for two-mode beating. At the lower limit of the rotation rate, 1 km s$^{-1}$, the rotational splitting of a trapped $\ell=1$ triplet is 0.00383 cd$^{-1}$. This corresponds to a beat period of 260 d for three-mode beating and 130 d for two-mode beating between the non-axisymmetric modes at the high inclination limit. It can be clearly seen that the timescales for amplitude modulation due to beating between the components of a $\ell=1$ multiplet are too short within the measured limits of rotation. Beating timescales, that would match the observed variability, require an improbable rotational velocity of $\leq$0.1 km s$^{-1}$.

We conclude that beating effects between components of a rotationally split triplet cannot explain the observed behavior. In principle, the observed long timescales could also be caused by a very close mode with a different spherical degree. The observational test relies critically on the observed phase shift at the time of minimum amplitude. Unfortunately, there is not enough data to carry out this test.

\subsection{Resonance effects}

Another cause for amplitude modulation is the resonant coupling of pulsation modes (e.g., Nowakowski 2005). Unstable pulsation modes are likely to be coupled with linearly stable modes of high spherical degrees. These modes may not be detected by photometric and spectroscopic methods due to cancellation effects across the stellar disk. However, to examine whether the amplitude variability of the $\ell = 1$ modes is due to resonance, the expected timescales of amplitude modulation can be compared to the observed variation.

Resonance effects show periodicities on timescales of the inverse linear amplitude growth rate of the unstable mode (e.g., Moskalik 1985). Table~\ref{tab:resonance} lists the expected timescales, $\tau_{\rm res}$, for all four observed $\ell = 1$ modes, according to a pulsation model for 44 Tau. 

\par
\begin{table}[htb!]
\caption{Predicted timescales for resonance effects.}
\label{tab:resonance}
\footnotesize
\begin{center}
\begin{tabular}{ccccc}
\noalign{\smallskip}
\hline\hline
\noalign{\smallskip}
 \multicolumn{2}{c}{Frequency [cd$^{-1}$]} &  $\tau_{\rm res} \approx 1/\gamma$ [yrs] & \multicolumn{2}{c}{$\tau_{\rm 1:1:1}$ [yrs]} \\
 &  &  &  V$_{\rm rot}$=1 km s$^{-1}$ &  V$_{\rm rot}$=5 km s$^{-1}$ \\
\noalign{\smallskip}
\hline
\noalign{\smallskip}
$f_2$ &  7.01 &  122.4 & 137 & 4.4 \\
$f_3$ &  9.12 &   90.5 & 55  & 2.6\\
$f_6$ &  9.56 &   16.1 & $\geq$150 & 17.1\\
$f_4$ & 11.52 &    1.3 & 45.6 & 1.8\\
\noalign{\smallskip}
\hline
\end{tabular}
\end{center}
\end{table}

The observed variability of the amplitude of the four $\ell$ = 1 modes occurs on a similar timescale, while the predicted modulation periods, $\tau_{\rm res}$, are very different. In particular, the observed variation in the amplitudes of $f_2$ and $f_3$  has a significantly shorter timescale than predicted. On the other hand, for $f_4$ the theoretical modulation timescale is too short.

We also investigated the possibility of a 1:1:1 resonance between the frequencies of an $\ell=1$ multiplet (see e.g. Buchler, Goupil \& Hansen 1997). The modulation period is $\tau_{\rm 1:1:1} \approx |1/(\nu_{+1} + \nu_{-1} - 2\nu_{0})|$ where the subscript denotes the $m$ value. For a model of 44~Tau with a rotational velocity at the lower limit of the measured rotation rate, the predicted values for $\tau_{\rm 1:1:1}$ have reasonable timescales with the exception of $f_6$ (see Table~\ref{tab:resonance}). For the upper limit of the rotation rate of 5.0 km s$^{-1}$, the computed timescales are too short for $f_2$, $f_3$, and $f_4$ and still too long to explain the observed variability of $f_6$.

Another difficulty concerns the nondetection of the other components of the rotational triplets. The components of the triplet are separated by $\sim$ 0.02 cd$^{-1}$ at the upper limit of the stellar rotation rate. We have only detected a
single possible case (9.58 cd$^{-1}$) with a small amplitude: regrettably, this mode does not have a mode identification. The missing detection of multiplet structures provides an argument against the 1:1:1 resonance hypothesis for rotation rates near our upper limit: due to the observed decrease in the amplitude of the observed mode, the amplitudes of the other components of the multiplet should already be large enough to be detectable. Consequently, while resonance between the components of a multiplet may explain the amplitude variability of a fraction of the $\ell$ = 1 modes, the 5-yr amplitude modulation of $f_6$ cannot be explained by the 1:1:1 resonance.

This section has shown that resonance may be responsible for the observed amplitude variability of specific modes. However, for definitive conclusions a longer time baseline for the observations would be required. Such data would also clarify whether the fascinating observation of a near-simultaneous decrease of the amplitudes of all the observed $\ell = 1$ modes is accidental.

\section{Impact of the new frequencies on the pulsation models of 44 Tau}

One of the main goals in this study was to search for new significant frequencies that extend the previously known frequency range of 6.34 -- 12.70 cd$^{-1}$ and to compare the new range to the predicted instability ranges for the two pulsation models given in Lenz et al. (2008). Mode instability is mainly determined by the conditions in the stellar envelope; e.g., the high-frequency limit in $\delta$ Scuti stars is very sensitive to the efficiency and treatment of envelope convection. In particular, near the terminal-age main sequence (TAMS), a small change in luminosity or temperature can imply very different stellar structures leading to different instability ranges. With a $\log g$ value of 3.6 $\pm$ 0.1 (Zima et al. 2007), 44~Tau represents such an ambiguous case.

The detection of a new frequency at 5.30 cd$^{-1}$ with an amplitude of 0.59 mmag poses a problem for the post-main sequence model. According to the results by Lenz et al. (2008), stability is predicted at such a frequency. We find that the low-frequency limit for instability cannot be shifted below this frequency by changing the input parameters (e.g., $\alpha_{\rm MLT}$) within reasonable limits. 
Consequently, there are two possibilities.

(i) The main sequence model represents the correct model. This assumption requires a high metallicity of Z = 0.03 to fit the identified radial modes. However, spectroscopic measurements show solar abundances in the photosphere (Zima et al. 2007). Moreover, this model cannot reproduce the measured fundamental parameters such as effective temperature.

(ii) The post-main sequence scenario, which matches the fundamental parameters, is correct, but the computed envelope structure of the model does not describe the real physical conditions in the star perfectly. Therefore, the predicted range of unstable modes is narrower than the observed frequency range. This problem also exists for other $\delta$ Scuti stars, such as FG Vir and for the two $\beta$ Cephei stars $\nu$~Eri and 12~Lac, as stated by Dziembowski \& Pamyatnykh (2008). These authors stress the necessity of an improvement in stellar opacity data by an opacity enhancement in the driving zone. Small opacity changes in the driving zone of $\delta$ Scuti stars may also make the mode at 5.30~cd$^{-1}$ unstable.

\section{Conclusions}

Two additional seasons of high-precision photometric data of the $\delta$ Scuti star 44 Tau were gathered with the Vienna Automatic Photoelectric Telescope in Arizona. A frequency analysis of the total data set including photometry from 2000 to 2006 led to detecting 49 pulsation frequencies, of which 15 are independent pulsation modes and 34 combination frequencies or harmonics.

We find that the amplitudes of the combination frequencies, $f_i + f_j$,  mirror the variations of the parent modes. The combination parameter, $\mu$, which relates the amplitudes of the combination frequencies to those of the parent modes, is found to be different for different radial and nonradial parents. The combination frequencies involving the radial fundamental mode have combination parameters, $\mu$, of 0.003, while those involving the first radial overtone can be as high as $\approx 0.009$.

The time-base of 5 years allows us to study amplitude variability in more detail. Strong amplitude variability from year to year was found for the $\ell$ = 1 modes, while the two radial modes have essentially constant amplitudes. Moreover, the amplitudes of all of the $\ell = 1$ modes decrease during most of the observed period. 
We examined several possible reasons for the amplitude variability of the $\ell$ = 1 modes:

(i) Precession of the pulsation axis: This causes variable amplitudes because the mode visibility depends on the inclination angle. While the amplitudes of the radial modes remain constant, the amplitudes of $\ell$ = 1 modes will change. However, both axisymmetric and non-axisymmetric modes decrease at the same time; this observation cannot be explained in this scenario because the visibility of $m = 0$ modes behaves opposite to their non-axisymmetric counterparts.

(ii) Beating of close frequencies: Due to the slow rotation of 44 Tau, beating between the components of a rotationally split $\ell=1$ triplet has been examined within the limits of the measured rotation rate. Generally, the predicted beating timescales are too short compared to the observed amplitude modulation. However, beating with hitherto unresolved, additional close frequencies (which are not a component of the same rotational triplet) could produce amplitude modulation at the observed timescales.

(iii) Resonance effects: The expected timescales for amplitude modulation in a 44 Tau model were computed. We find that only for a fraction of the variable $\ell = 1$ modes could the predicted variability timescales match the observed modulation periods.

None of these three explanations can explain the observed amplitude variability completely. However, a combination of all these effects cannot be excluded. Even more data would be required to clarify this point.

The newly found gravity mode at 5.30 cd$^{-1}$ extends the previously known frequency range to lower frequencies. This detection agrees better with the predicted instability ranges for main sequence models. However, our main sequence models cannot fit the measured fundamental parameters such as effective temperature. Our post-main sequence models provide a good fit of the fundamental parameters but predict stability at 5.30 cd$^{-1}$. The problem with the present models underestimating the observed $\delta$ Scuti star instability is common to many $\delta$ Scuti stars and may be related to still incomplete opacity data.

\begin{acknowledgements}

It is a pleasure to thank Alosha Pamyatnykh and Pawel Moskalik for helpful discussions. This investigation
has been supported by the Austrian Fonds zur F\"orderung der wissenschaftlichen Forschung.

\end{acknowledgements}

\end{document}